\documentstyle[preprint,aps,psfig]{revtex}

\begin{document}

\draft
\preprint{\parbox{5cm}{ADP-95-31/T185 \\
                       TUM/T39-95-8 \\ \\}}
\title{$Q^2$ Dependence of Nuclear Shadowing}

\author{W.Melnitchouk}
\address{Physik Department,
         Technische Universit\"at M\"unchen,
         D-85747 Garching, Germany.}
\author{A.W.Thomas}
\address{Department of Physics and Mathematical Physics,
         University of Adelaide, 5005, Australia}

\maketitle

\begin{abstract}
We re-examine the predictions of a two-phase model of shadowing
in nuclear deep-inelastic scattering in light of new NMC data
on the $x$, $A$ and $Q^2$ dependence of the ratios of
structure functions.
The model, which combines vector meson dominance at low
$Q^2$ with diffractive Pomeron exchange for large $Q^2$,
agrees with the observed small, but non-zero, slopes in $\log Q^2$,
which indicate the importance of higher twist shadowing effects
in the transition region, $0.1 \alt Q^2 \alt 1$ GeV$^2$.
We note also that the latest E665 data on the deuteron to proton
ratio suggests the presence of a small amount of shadowing in the
deuteron.
\end{abstract}

\vspace*{0.5cm}

\pacs{PACS numbers: 13.60.Hb; 12.40.Vv; 12.40.Gg.
      \vspace*{1cm} \\
      To appear in Phys.Rev.C }


There has recently been a great deal of progress in the experimental
study of the $A$ dependence of nuclear structure functions in the
small-$x$ region where nuclear shadowing occurs.
At CERN the New Muon Collaboration (NMC) has been able to extract
the slopes in the four-momentum transfer squared, $Q^2$, of the ratios
of structure functions of a variety of nuclei for $x = Q^2/2M\nu$
down to $\sim$ 0.001 \cite{N_91,N_RE,N_LIC}, where $\nu$ is the energy
transfer to the target and $M$ the nucleon mass.
Meanwhile, E665 at Fermilab has extended previous measurements
on the deuteron and several heavier nuclei to $x$ as low as
$x \sim 10^{-5}$ \cite{E_D,E_D93,E_A,E_XE}.

These new measurements provide a tremendous challenge to theoretical
models of nuclear deep-inelastic scattering.
Not only do they cover five orders of magnitude in $x$, but the
momentum transfer squared also varies from tens of GeV$^2$ to
values as low as 0.05 GeV$^2$.
For the latter one is (to say the least) hard pressed to justify
a parton model description.
Clearly here one is dealing with a transition region between
those regimes where hadronic and explicit quark treatments
are appropriate.
While it is unlikely that any single theoretical approach will be
totally reliable over the entire range of $x$ and $Q^2$, it is
important that all existing models be put to the test with a view to
understanding the nature of the transition between high
and low $Q^2$.

Our aim is to extend earlier calculations of shadowing in deuterium
\cite{MTD} and heavier nuclei \cite{MTA} to compare with these new
data.
The approach taken is a two-phase model, similar to the work of
Kwiecinski and Badelek \cite{KWBD,KW,BK}.
At high virtuality the interaction of the virtual photon with the
nucleus is most efficiently parametrized in terms of diffractive
scattering through the double and triple Pomeron.
On the other hand, at low virtuality it is most natural to apply
a vector meson dominance (VMD) model, in which the virtual photon
interacts with the nucleons via its hadronic structure,
namely the $\rho^0$, $\omega$ and $\phi$ mesons.
The latter contribution vanishes at sufficiently high $Q^2$, but
in the region of interest here it is in fact responsible for the
majority of the $Q^2$ variation.

For convenience we shall summarize the key ingredients of the
calculation, beginning with the diffractive component.
For this we use Pomeron ($I\!\!P$) exchange between the projectile
and two or more constituent nucleons to model the interaction of
partons from different nucleons within the nucleus.
Assuming factorization of the diffractive cross section \cite{INGS},
the shadowing correction (per nucleon) to the nuclear structure
function $F_2^A$ from $I\!\!P$-exchange is written as a convolution
of the Pomeron structure function, $F_2^{I\!P}$, with a distribution
function (``flux factor''), $f_{I\!P/A}$, describing the number
density of exchanged Pomerons:
\begin{eqnarray}
\label{dFAP}
\delta^{(I\!P)} F_2^A(x,Q^2)
&=& {1 \over A} \int_{y_{min}}^A\ dy\
f_{I\!P/A}(y)\ F_2^{I\!P}(x_{I\!P},Q^2),
\end{eqnarray}
where $y = x (1+M_X^2/Q^2)$ is the light-cone momentum fraction
carried by the Pomeron ($M_X$ is the mass of the diffractive
hadronic debris), and $x_{I\!P} = x/y$ is the momentum fraction
of the Pomeron carried by the struck quark in the Pomeron.
The $y$ dependence of $f_{I\!P/A}(y)$ can be calculated within
Regge theory \cite{MTD,MTA,KWBD,KW,BK,DOLA}
(for a survey of other definitions of the ``flux factor'' see
Ref.\cite{PRY}).
Within experimental errors, the factorization hypothesis, as well as
the $y$ dependence of $f_{I\!P/A}(y)$ \cite{MTD,MTA,KWBD,KW,BK},
are consistent with the results obtained recently by the H1 \cite{H1}
and ZEUS \cite{ZEUS} Collaborations at HERA from observations of
large rapidity gap events in diffractive $ep$ scattering.
These data also confirm previous findings by the UA8 Collaboration
\cite{UA8} that the Pomeron structure function contains both a hard
and a soft component:\
$ F_2^{I\!P}(x_{I\!P},Q^2)
= F_2^{I\!P ({\rm hard})}(x_{I\!P},Q^2)
+ F_2^{I\!P ({\rm soft})}(x_{I\!P},Q^2)$.
The hard component of $F_2^{I\!P}$ is generated from
an explicit $q\bar q$ component of the Pomeron, and has
an $x_{I\!P}$ dependence given by\
$x_{I\!P} (1-x_{I\!P})$ \cite{DOLA,NZ}, in agreement
with the recent diffractive data \cite{H1,ZEUS,UA8}.
The soft part, which is driven at small $x_{I\!P}$ by the
triple-Pomeron interaction \cite{KWBD,NZ}, has a sea quark-like
$x_{I\!P}$ dependence\footnote{
In Ref.\cite{NZ} the soft component is associated
the $q\bar q g$ Fock component of the $\gamma^*$ wave function.},
with normalization fixed by the triple-Pomeron coupling constant
\cite{GOUL}.

The dependence of $F_2^{I\!P}$ on $Q^2$ at large $Q^2$ arises
from radiative corrections to the parton distributions in the
Pomeron \cite{KW,CHPWW}, which leads to a weak, logarithmic,
$Q^2$ dependence for the shadowing correction
$\delta^{(I\!P)} F_2^A$.
Alone, the Pomeron contribution to shadowing would give a
structure function ratio $F_2^A/F_2^D$ that would be almost
flat for $Q^2 \agt 2$ GeV$^2$.
The low-$Q^2$ extrapolation of the $q\bar q$ component is
parametrized by applying a factor $Q^2/(Q^2 + Q_0^2)$, where
$Q_0^2 \approx 0.485$ GeV$^2$ \cite{DLQ} may be interpreted
as the inverse size of partons inside the virtual photon.
For the nucleon sea quark densities relevant for
$F_2^{I\!P ({\rm soft})}$
we use the recent parametrization from Ref.\cite{DLQ}, which includes
a low-$Q^2$ limit consistent with the real photon data, in which
case the total Pomeron contribution
$\delta^{(I\!P)} F_2^A \rightarrow 0$ as $Q^2 \rightarrow 0$.

To adequately describe shadowing for small $Q^2$ requires one to
use a higher-twist mechanism, such as vector meson dominance.
VMD is empirically based on the observation that some aspects
of the interaction of photons with hadronic systems resemble
purely hadronic interactions \cite{BAUER,SCHSJ}.
In terms of QCD this is understood in terms of a coupling of the
photon to a correlated $q\bar q$ pair of low invariant mass,
which may be approximated as a virtual vector meson.
One can then estimate the amount of shadowing in terms of the
multiple scattering of the vector meson using Glauber theory.
The corresponding correction (per nucleon) to the nuclear
structure function is:
\begin{eqnarray}
\label{dFAV}
\delta^{(V)} F_2^A(x,Q^2)
&=& {1 \over A} { Q^2 \over \pi }
\sum_V
{ M_V^4\ \delta\sigma_{VA} \over f_V^2 (Q^2 + M_V^2)^2 },
\end{eqnarray}
where $\delta \sigma_{VA}$ is the shadowing correction to the
vector meson---nucleus cross section, calculated in Ref.\cite{MTA},
$f_V$ is the photon---vector meson coupling strength \cite{BAUER}
(see also \cite{CSB}), and $M_V$ is the vector meson mass.
In practice, only the lowest mass vector mesons
($V = \rho^0, \omega, \phi$) are important at low $Q^2$.
(Inclusion of higher mass vector mesons, including continuum
contributions, leads to so-called generalized vector meson dominance
models \cite{GVMD}.)
Usually one omits non-diagonal vector meson transitions
($VN \rightarrow V'N$), as these are not expected to be large.
The vector meson propagators in Eq.(\ref{dFAV}) lead to a strong
$Q^2$ dependence of $\delta^{(V)} F_2^A$, which peaks at
$Q^2 \sim 1$ GeV$^2$.
For $Q^2 \rightarrow 0$ and fixed $x$, $\delta^{(V)} F_2^A$
disappears due to the vanishing of the total $F_2^A$.
Furthermore, since this is a higher twist effect, shadowing in the
VMD model dies off quite rapidly between $Q^2 \sim 1$ and 10 GeV$^2$,
so that for $Q^2 \agt 10$ GeV$^2$ it is almost negligible ---
leaving only the diffractive term, $\delta^{(I\!P)} F_2^A$.
(Note that at fixed $\nu$, for decreasing $Q^2$ the ratio
$F_2^A/F_2^D$ approaches the photoproduction limit.)

While the asymptotic $Q^2$ behavior of nuclear shadowing seems clear,
there is still considerable interest in the transition region where
the high- and low-$Q^2$ descriptions merge.
In practice, this occurs for $Q^2$ between about 0.5 and 5 GeV$^2$,
which is precisely the region where most of the recent NMC data on
$C$, $Ca$ and other nuclei have been taken \cite{N_91,N_RE,N_LIC}.

For light nuclei, such as $C$, the dominant mechanism for nuclear
shadowing involves the double scattering of the projectile from
two nucleons.
Higher order terms (multiple rescattering) in the Glauber expansion
attenuate the incident flux of vector mesons (or of the hadronic
state $X$ for the diffractive component \cite{KW}) as they traverse
the nucleus, which will be progressively more important as $A$
increases \cite{MTA}.
For the VMD component, the magnitude of the attenuation is
determined by the mean free path, $L_V$, of the vector meson in the
nucleus, $L_V = (\rho\ \sigma_{VN})^{-1}$, where $\sigma_{VN}$ is
the total $VN$ cross section \cite{DLSIG}.
If the mean free path $L_X = (\rho\ \sigma_{XN})^{-1}$ of the hadronic
state $X$ is independent of the mass $M_X$ \cite{KW}, one may take
the reabsorption cross section
$\sigma_{XN} \sim \sigma_{VN} \sim 20-30$ mb,
although in the $Q^2$ range covered by the NMC data the $F_2^A/F_2^D$
ratios are not very sensitive to the precise value of $\sigma_{XN}$.
For the single particle density $\rho$ in heavy nuclei ($A \agt 16$)
we use the Woods-Saxon (or Fermi) density, while for light nuclei
($A \alt 16$) the harmonic oscillator (shell model) form is more
appropriate \cite{GS}.
Short range correlations are included through a Fermi gas correlation
function which puts a ``hole'' in the two-body density approximately
0.5~fm wide at 1/2 maximum density \cite{MTA}.
The inclusion of correlations has the effect of decreasing slightly
the amount of shadowing (i.e. increasing the ratio $F_2^A/F_2^D$)
at low $Q^2$.

Having outlined the essential features of the model, we now turn
to a detailed comparison with the data.
In Fig.1 we plot the ratio of the structure functions $F_2$
(normalized to one nucleon) for $C$ and $D$ as a function of $Q^2$,
for various values of $x$ ranging from $x = 0.0003$ to $x = 0.055$.
The data represent the complete sample taken by the NMC for $C$
nuclei.
The overall agreement between the model calculation and the
data is clearly excellent.
In particular, the observed $Q^2$ dependence of the ratios
is certainly compatible with that indicated by the NMC data.
At large $Q^2$ ($Q^2 \agt 10$ GeV$^2$), the calculated
curves are almost constant with $Q^2$, as would be expected
from a partonic (scaling) mechanism \cite{NZ,QIU,BL,KUM,KP}.
However, in the smallest $x$ bins the $Q^2$ values reach
as low as $Q^2 \approx 0.05$ GeV$^2$.
Clearly this region of $Q^2$ is inaccessible to any model
involving only a partonic mechanism, and it is essential
here to invoke a non-scaling mechanism such as the VMD model.

It is common practice in many data analyses to extract slopes
in $\log Q^2$ by performing simple straight-line fits to the data
--- such as $F_2^A / F_2^D = a + b \log Q^2$.
To illustrate a potential difficulty with using such extracted
slopes to discriminate between different models of shadowing,
we plot in Fig.1 (for $x = 0.0125$) the slope
that one would obtain by making a linear fit to the
$\log Q^2$ variation of $F_2^C/F_2^D$.
Evidently the slope thus obtained is negative, while the calculated
curves have a positive slope.
Within the present degree of accuracy, the $C/D$ data are equally
compatible with a positive or negative slope, and the trend of the
data as a function of $Q^2$ clearly favors the positive value obtained
in our calculation.

In Fig.2 we show the structure function ratio for $Ca/D$,
for $x$ between $x = 0.0085$ and 0.07.
Again, within the error bars, the calculated $x$ and $Q^2$
dependence is in excellent agreement with the data.
The apparent small decrease in the ratio for the largest
$Q^2$ points could be due to a small nuclear dependence
of the ratio of the longitudinal to transverse virtual
photoabsorption cross sections \cite{MIL}, which is assumed
to be zero in the data extraction.
However, the statistics on the data do not allow a definitive
statement about this effect.

In future the NMC will also produce data on the $Q^2$ dependence
of the $Sn$ to $C$ ratio \cite{MUE}.
In Fig.3(a) we show the predictions for the structure function
ratio as a function of $Q^2$, for $x = 0.0125$ (lowest curve),
0.0175, 0.025, 0.035, 0.045 and 0.055 (highest curve).
Because of the expected higher accuracy of these data \cite{MUE},
linear fits to the $\log Q^2$ dependence should result
in more reliable extracted slopes.
In Fig.3(b) we show the slopes
$b = d \left(F_2^{Sn}/F_2^C\right) / d\log Q^2$,
extracted from the curves in Fig.3(a), as a function of $x$.
The solid line is the slope for the VMD and $I\!\!P$-exchange
mechanisms, while the dashed line represents the Pomeron
contribution only.
We would like to stress that essentially zero slopes
would be obtained in models where only a partonic
mechanism would be utilized \cite{NZ,QIU,BL,KUM}.
Observation of non-zero slopes would clearly support the
hypothesis that the intermediate-$Q^2$ region is dominated by
non-scaling, higher-twist effects, having a non-trivial $Q^2$
dependence.

Data for other nuclei have also been taken recently by the NMC.
The $A$ dependence of the structure function ratio of deuterium,
$Li$, $Be$, $Al$, $Ca$, $Fe$ and $Sn$ to carbon has been
parametrized as $F_2^A(x)/F_2^C(x) \propto A^{\alpha(x)}$.
In Fig.4 we show the slope
$\alpha(x) = d \left( F_2^A/F_2^C \right) / d \log A$
as a function of $x$.
The agreement at the low $x$ values is clearly very good.
At larger values of $x$ ($x \agt 0.07$) the data go above unity,
and tend to lie slightly above the calculated curve.
This is a reflection of the fact that the present model
has no mechanism for antishadowing built in.

Having obtained excellent fits to the nuclear data, we can be
reasonably confident that applying the same model to the deuteron
\cite{MTD,BK,ZOL,KH} will yield reliable results.
Indeed, here one has fewer model parameters to deal with,
since all of the shadowing is generated through double
scattering alone.
As was observed in Refs.\cite{MTD,BK}, the largest uncertainty
in the calculation of the shadowing in deuterium is the
deuteron wave function.
Nevertheless, the presence of shadowing in the deuteron would
be confirmed through observation of a deviation from unity in
the $D/p$ structure function ratio in the kinematic region
where Regge theory is expected to be valid.
Although the exact value of $x$ below which the proton and
(free) neutron structure functions become equivalent is not
known, it is expected that at low enough $x$,
$F_2^p \rightarrow F_2^n$, in which case
$F_2^D/F_2^p \rightarrow 1 + \delta F_2^D/F_2^p$.
In Fig.5 we show the data at very low $x$ taken by the E665
Collaboration \cite{E_D}, as well as the earlier NMC data
at larger $x$ \cite{N_D}.
The calculated ratio with a small shadowing correction is shown
by the solid curve, while the result for the case of no shadowing
is indicated by the dashed curve.
The data clearly favor the shadowing scenario.
As a fraction of $F_2^D$, the shadowing correction amounts to
about 1.5\% at $x = 10^{-2}$ up to about 3\% for $x \alt 10^{-5}$.

In summary, we have seen that the latest data from the NMC
and E665 indicate that nuclear shadowing in the low- and
intermediate-$Q^2$ regions is controlled by the dynamics
of a higher-twist, vector meson dominance mechanism.
At larger $Q^2$ this component disappears, leaving behind a
scaling component which is understood to arise from diffractive
scattering from the Pomeron component of the nucleon, and which
agrees with the approximate $Q^2$ independence of the data at
large $Q^2$.
Models based solely on partonic mechanisms can therefore provide
only limited insight into the physics of nuclear shadowing.

\acknowledgements

We would like to thank A.Br\"ull and A.M\"ucklich for providing
the NMC structure function ratios, and P.Spentzouris
for sending the E665 $D/p$ data points.
We also acknowledge informative discussions with
J.Milana, G.Piller, W.Ratzka, T.Sloan and W.Weise.


\begin{figure}
\centering{\ \psfig{figure=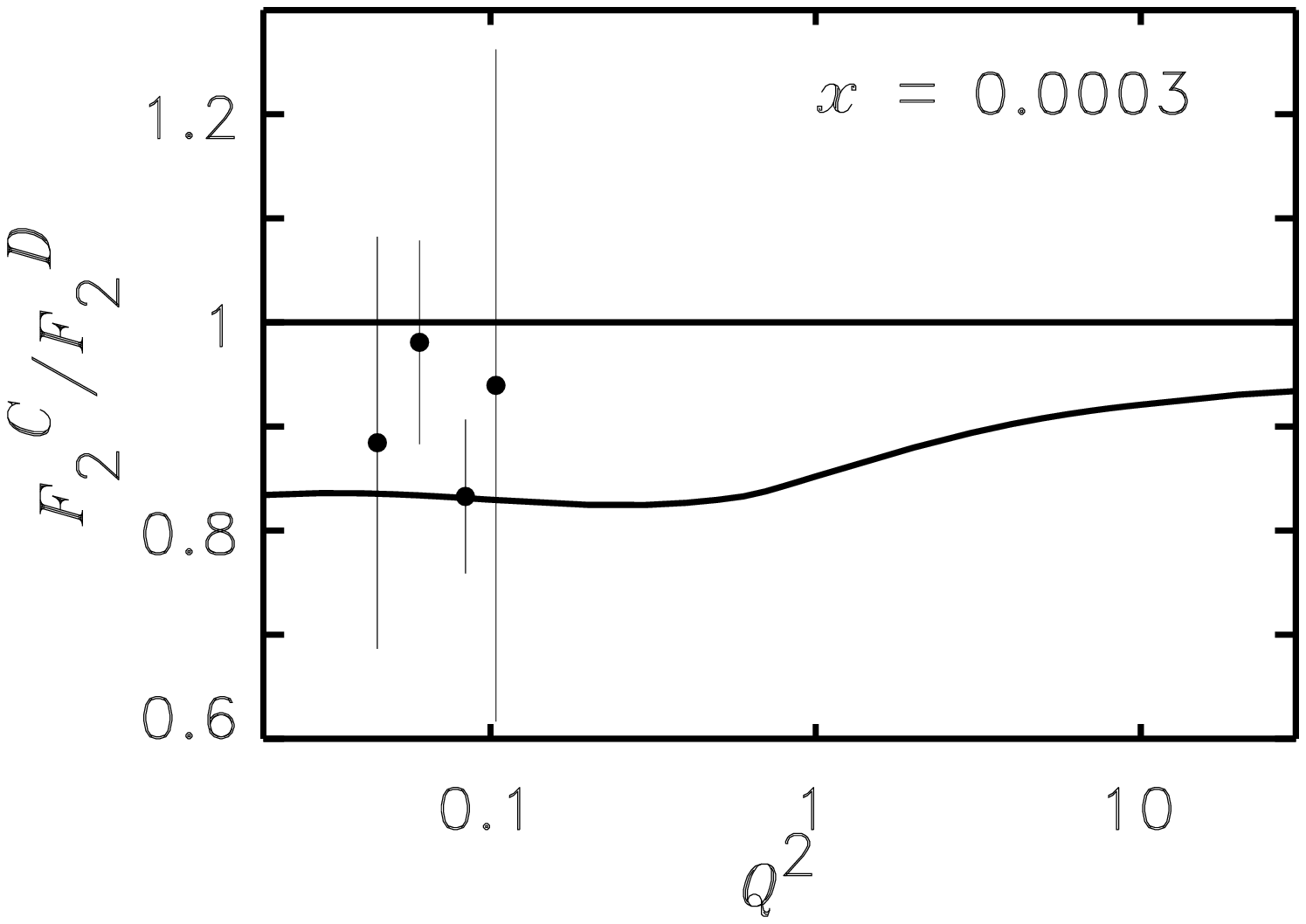,height=7cm}}
\centering{\ \psfig{figure=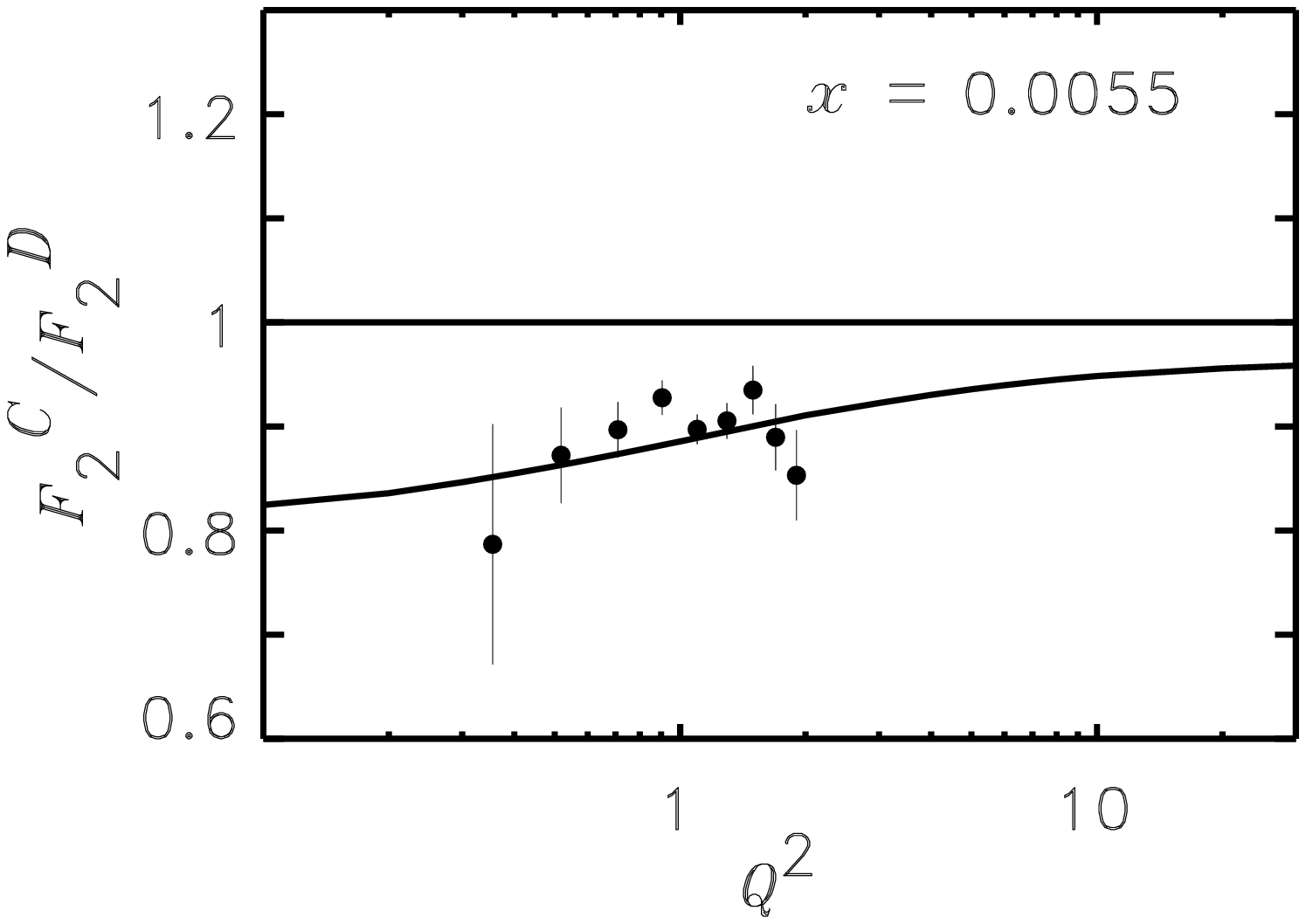,height=7cm}}
\centering{\ \psfig{figure=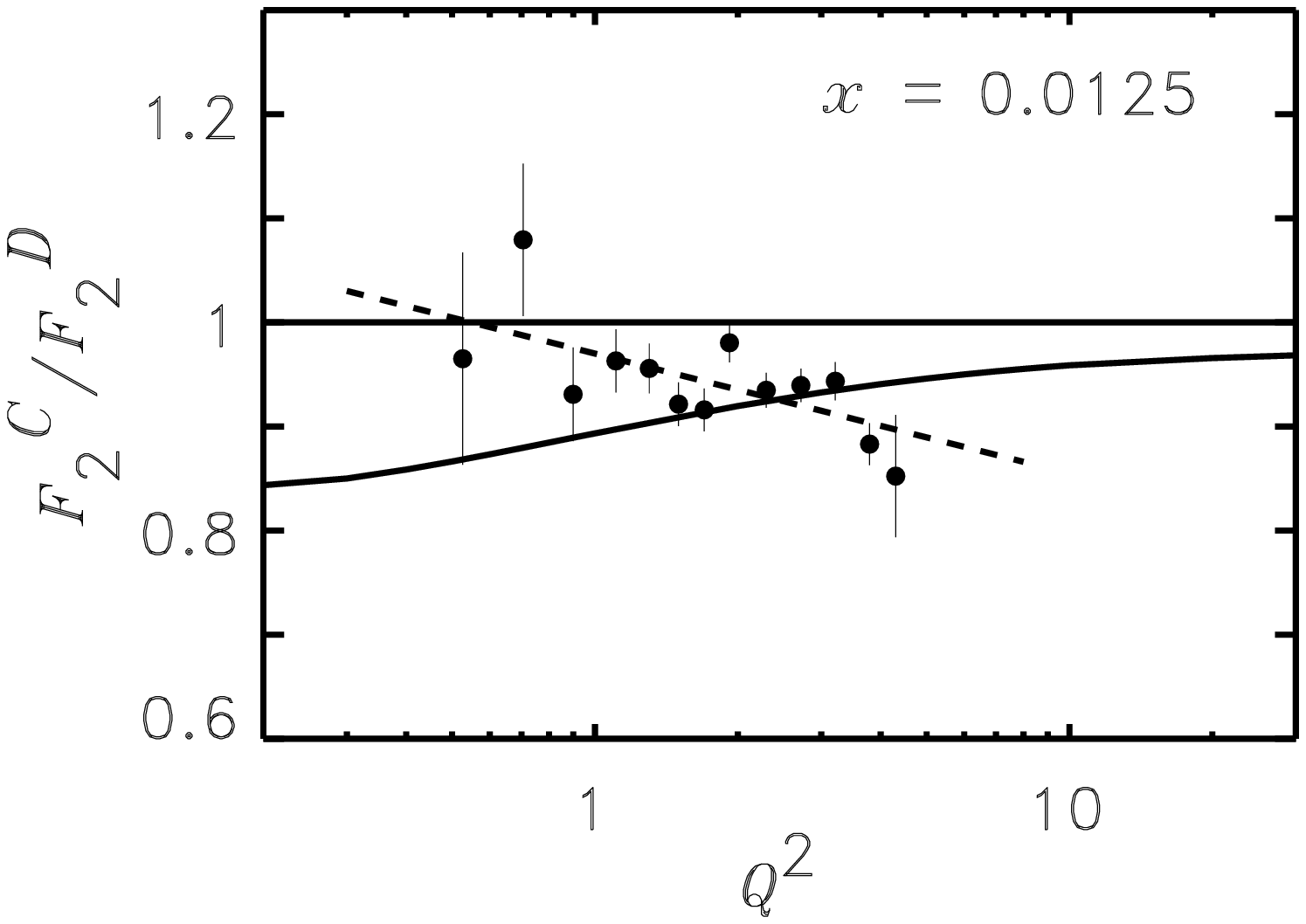,height=7cm}}
\newpage
\centering{\ \psfig{figure=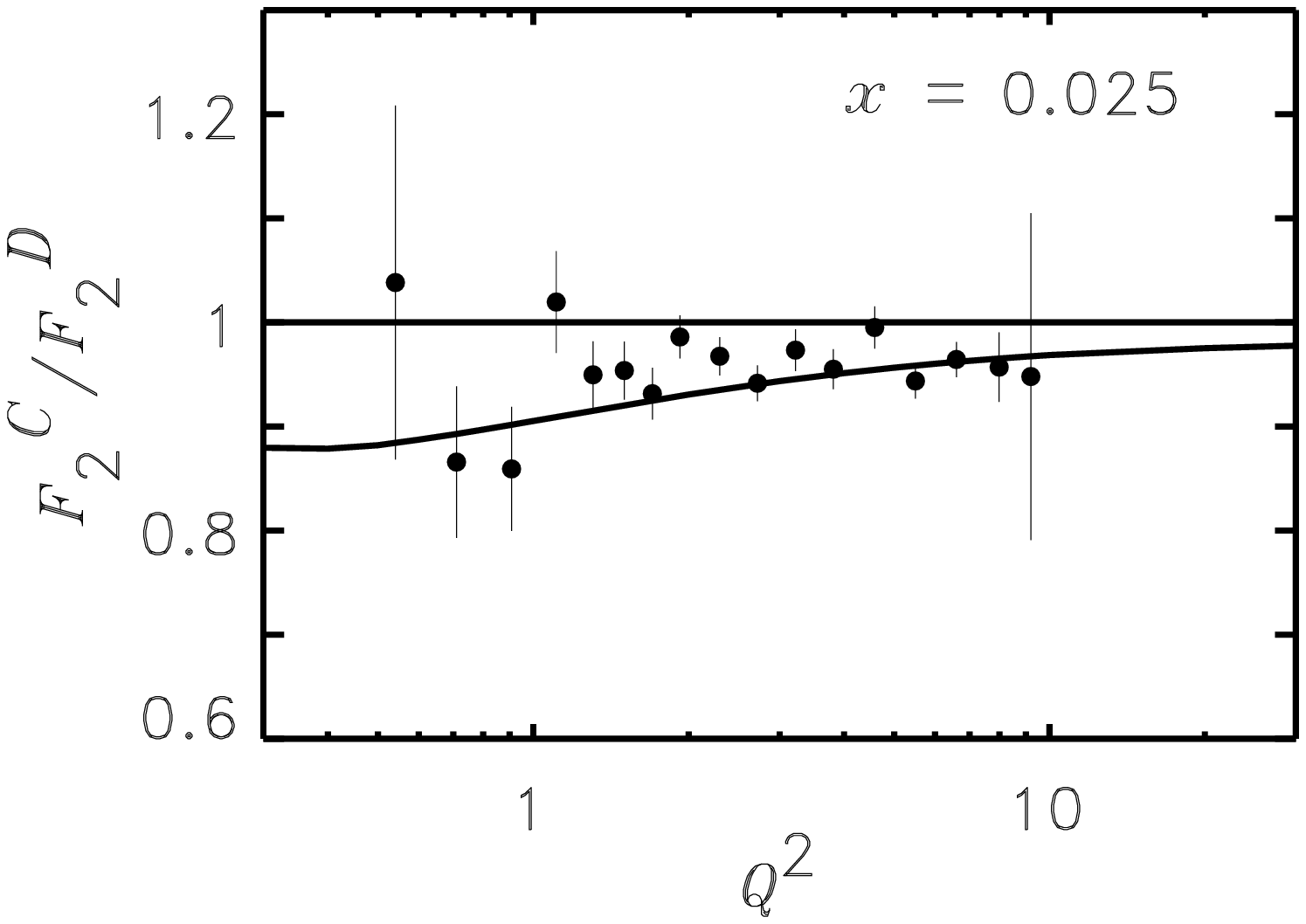,height=7cm}}
\centering{\ \psfig{figure=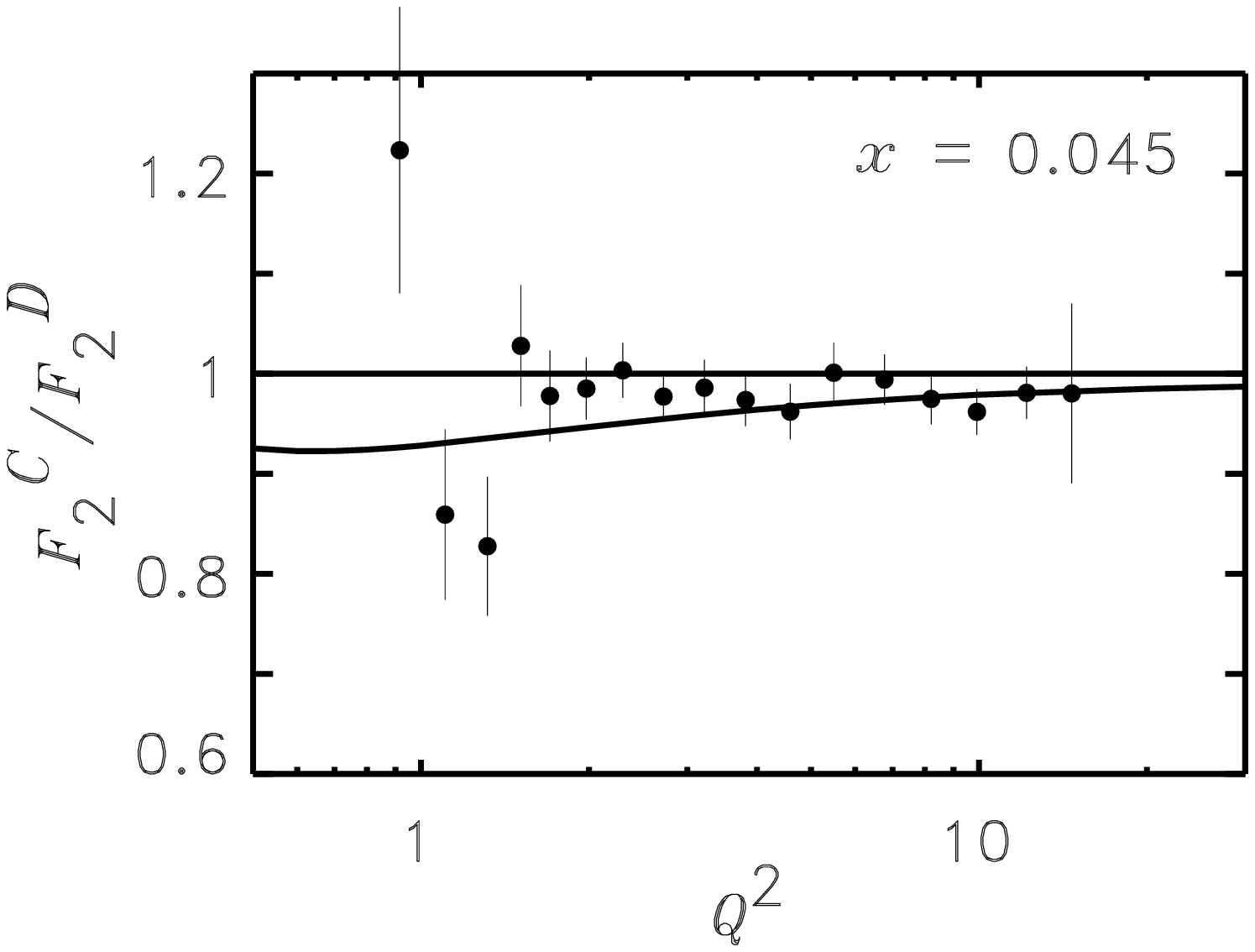,height=7cm}}
\centering{\ \psfig{figure=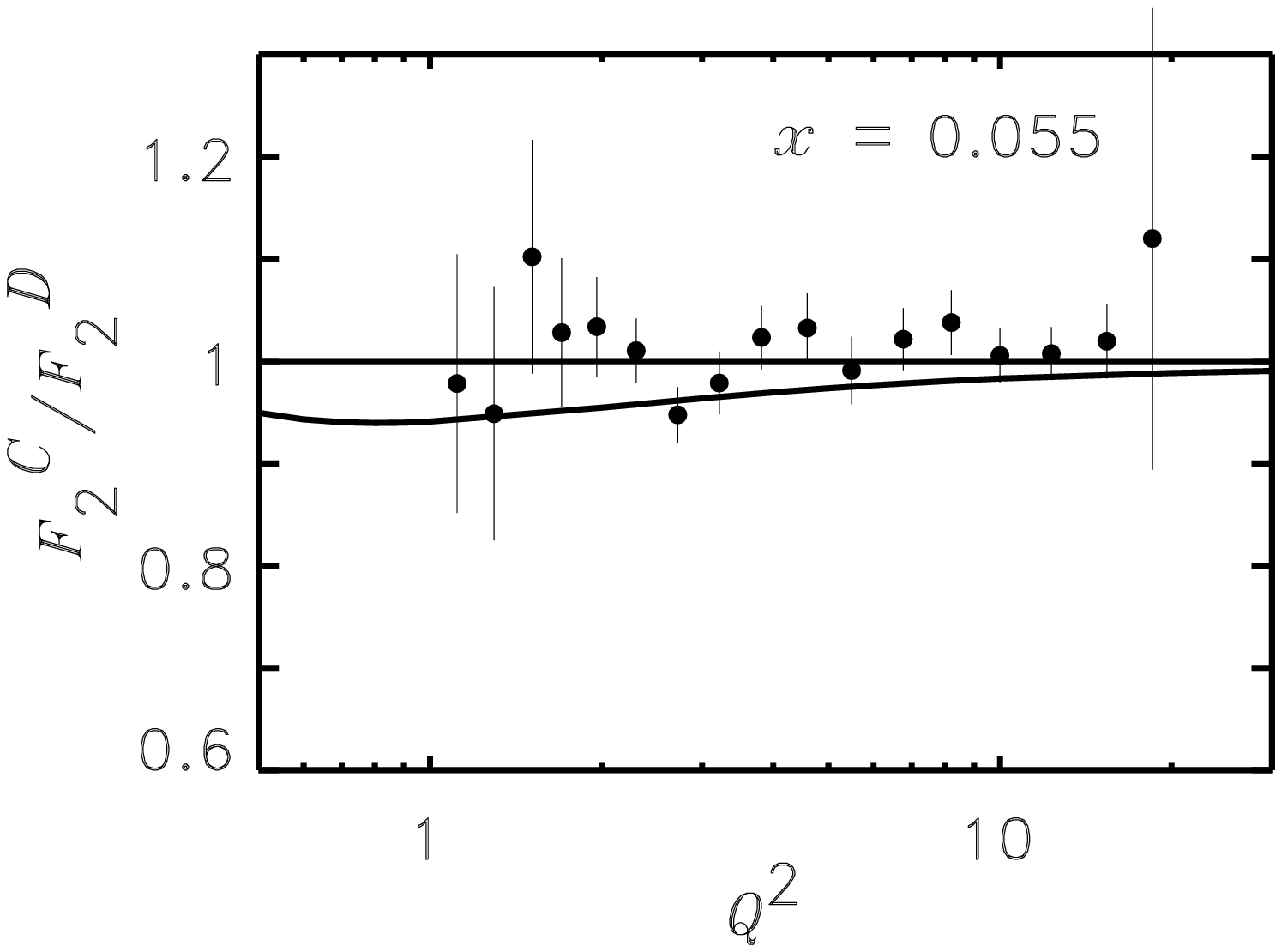,height=7cm}}
\caption{$Q^2$ variation of the $C/D$ structure function ratio,
        for various values of $x$
        ($x$ = 0.0003, 0.0055, 0.0125, 0.025, 0.045 and 0.055),
        compared with the complete NMC data \protect\cite{N_RE,N_LIC}.
        Also shown for $x=0.0125$ is a linear fit in $\log Q^2$
        to the data.}
\end{figure}

\begin{figure}
\centering{\ \psfig{figure=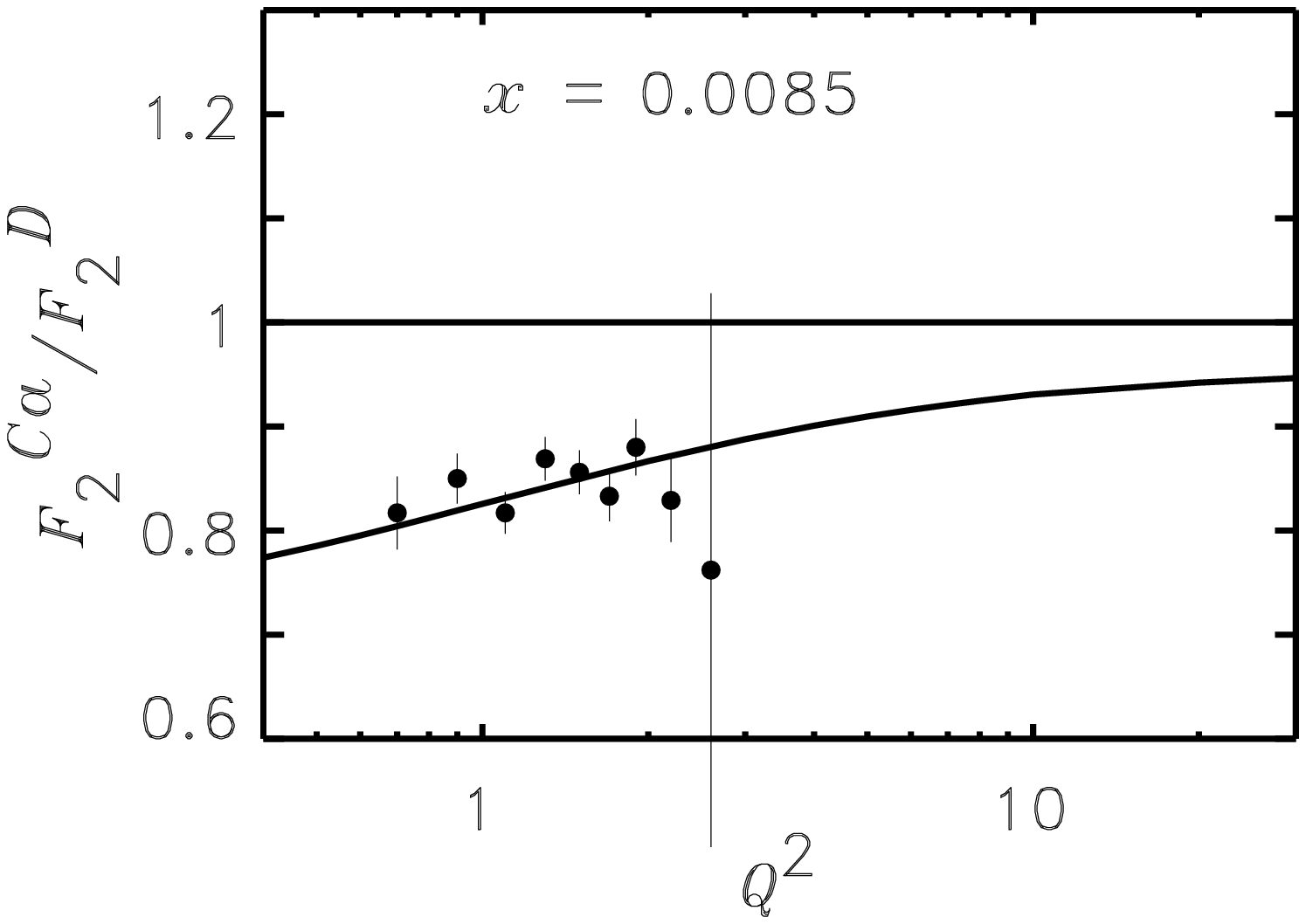,height=7cm}}
\centering{\ \psfig{figure=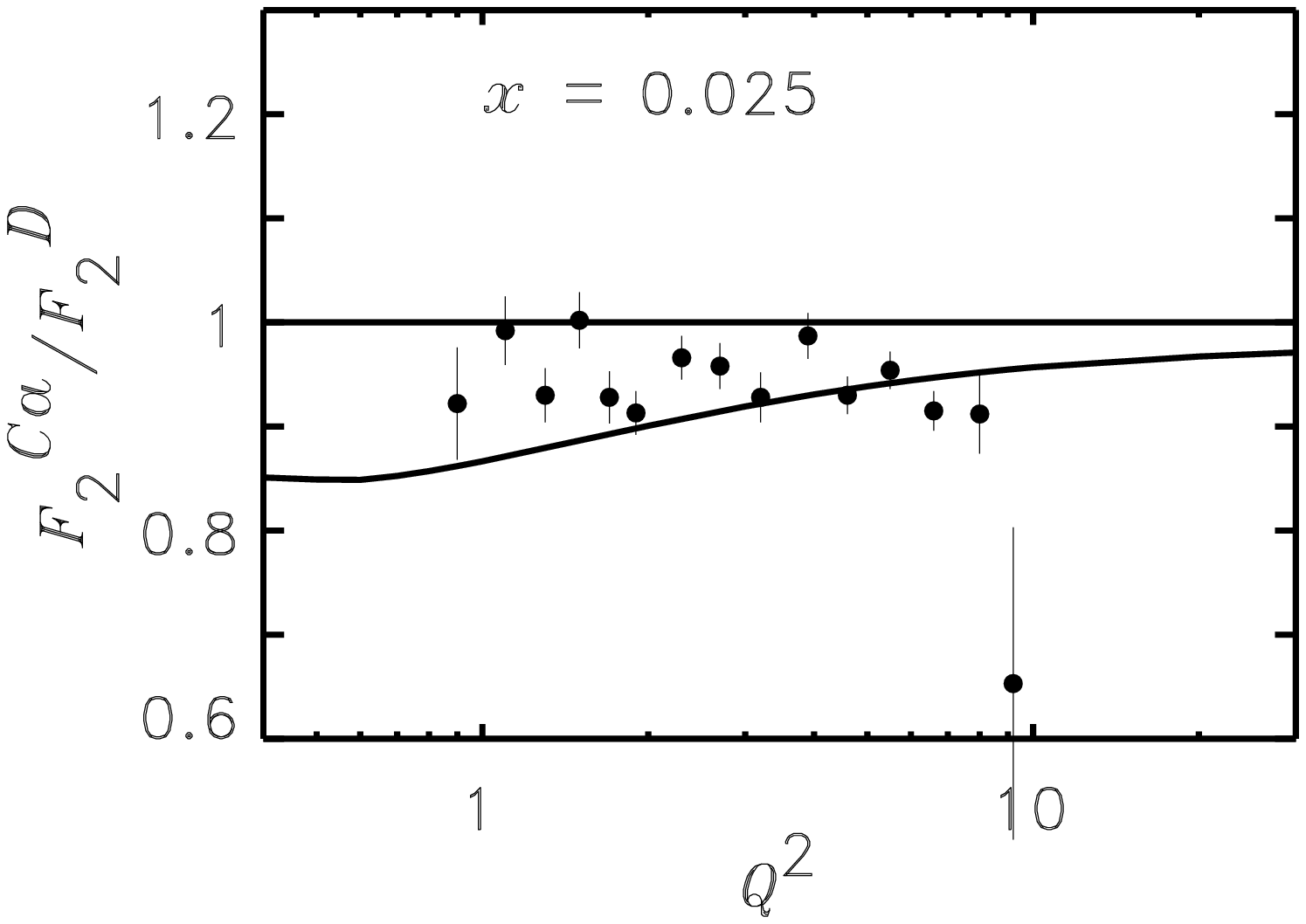,height=7cm}}
\centering{\ \psfig{figure=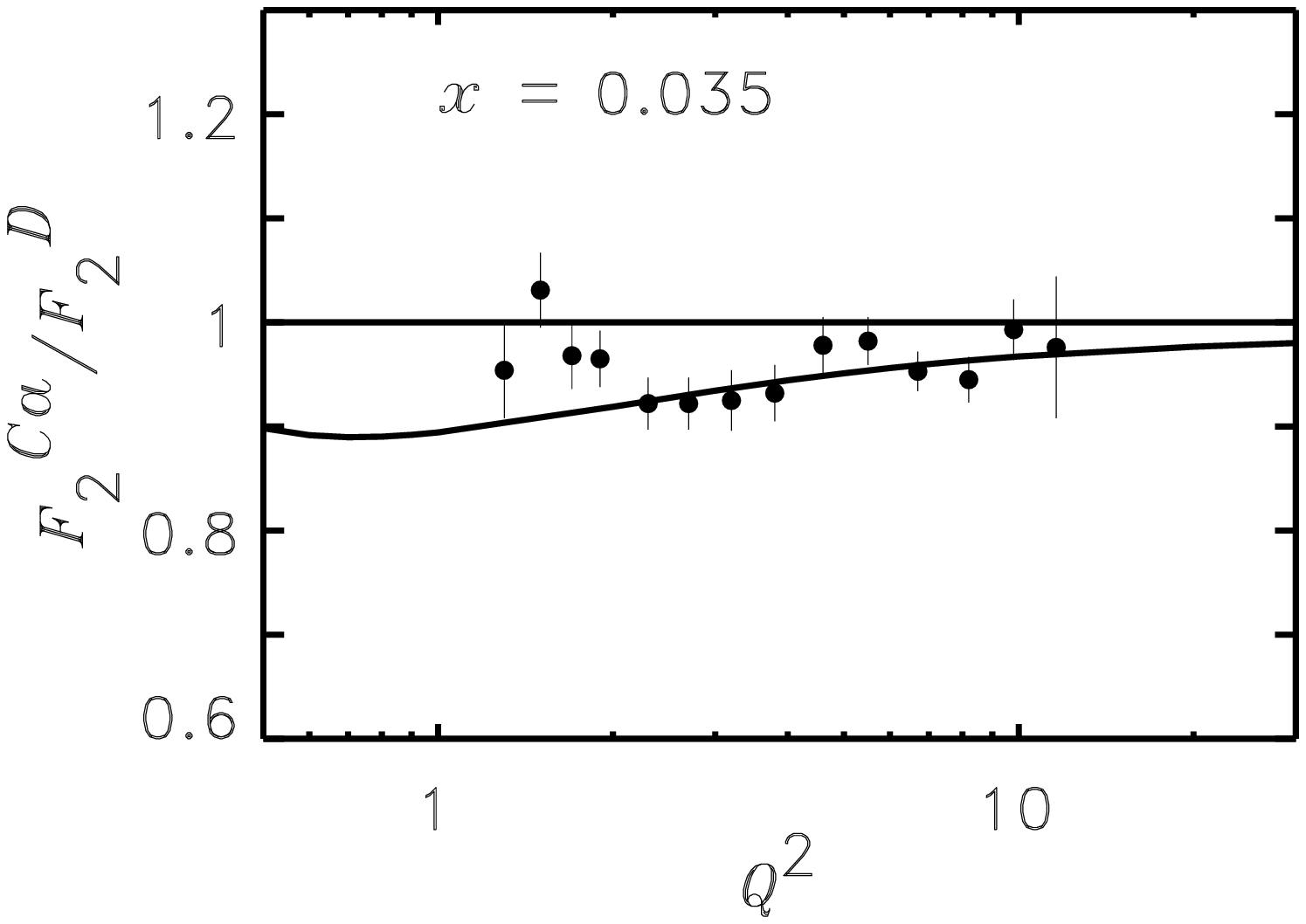,height=7cm}}
\newpage
\centering{\ \psfig{figure=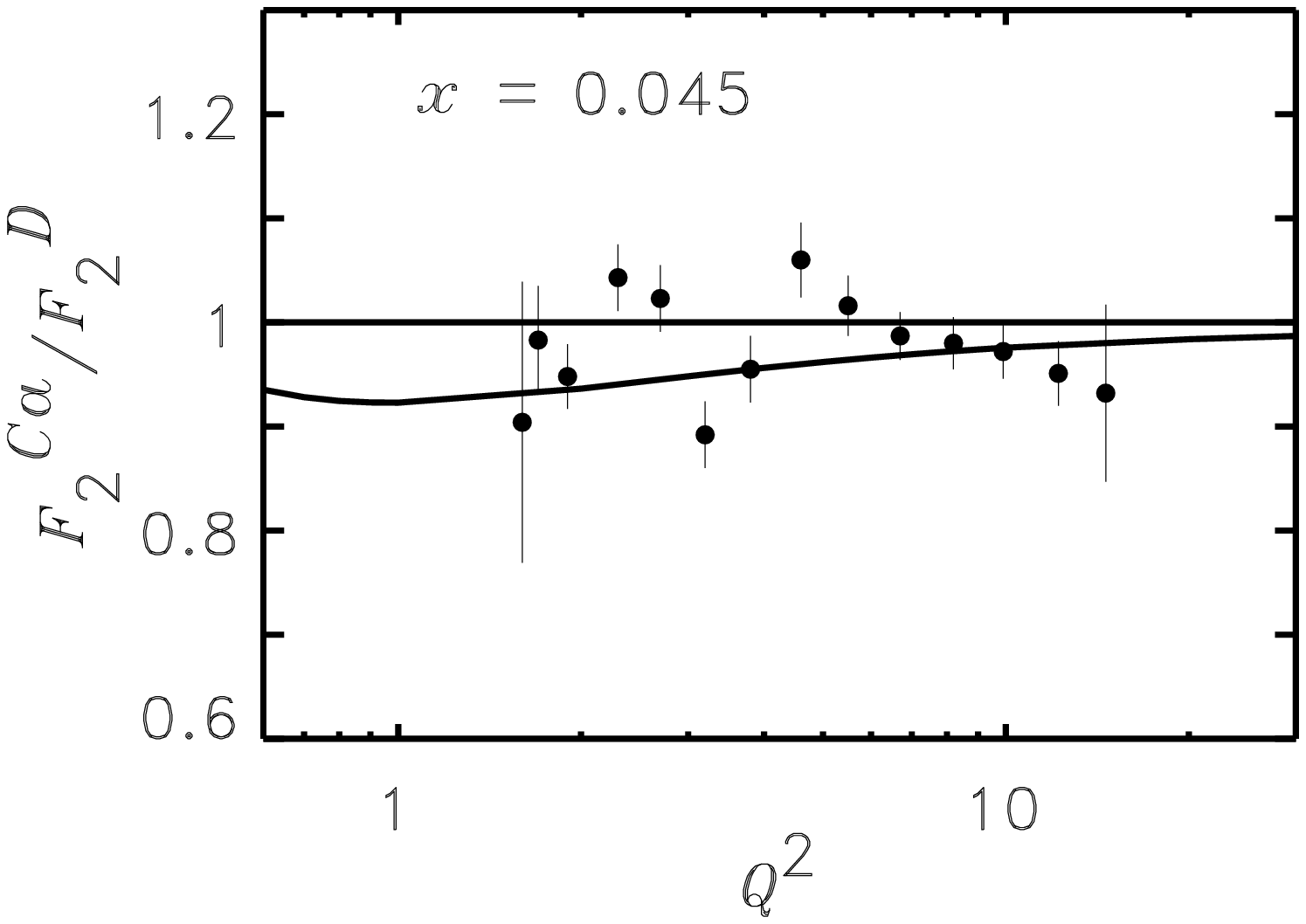,height=7cm}}
\centering{\ \psfig{figure=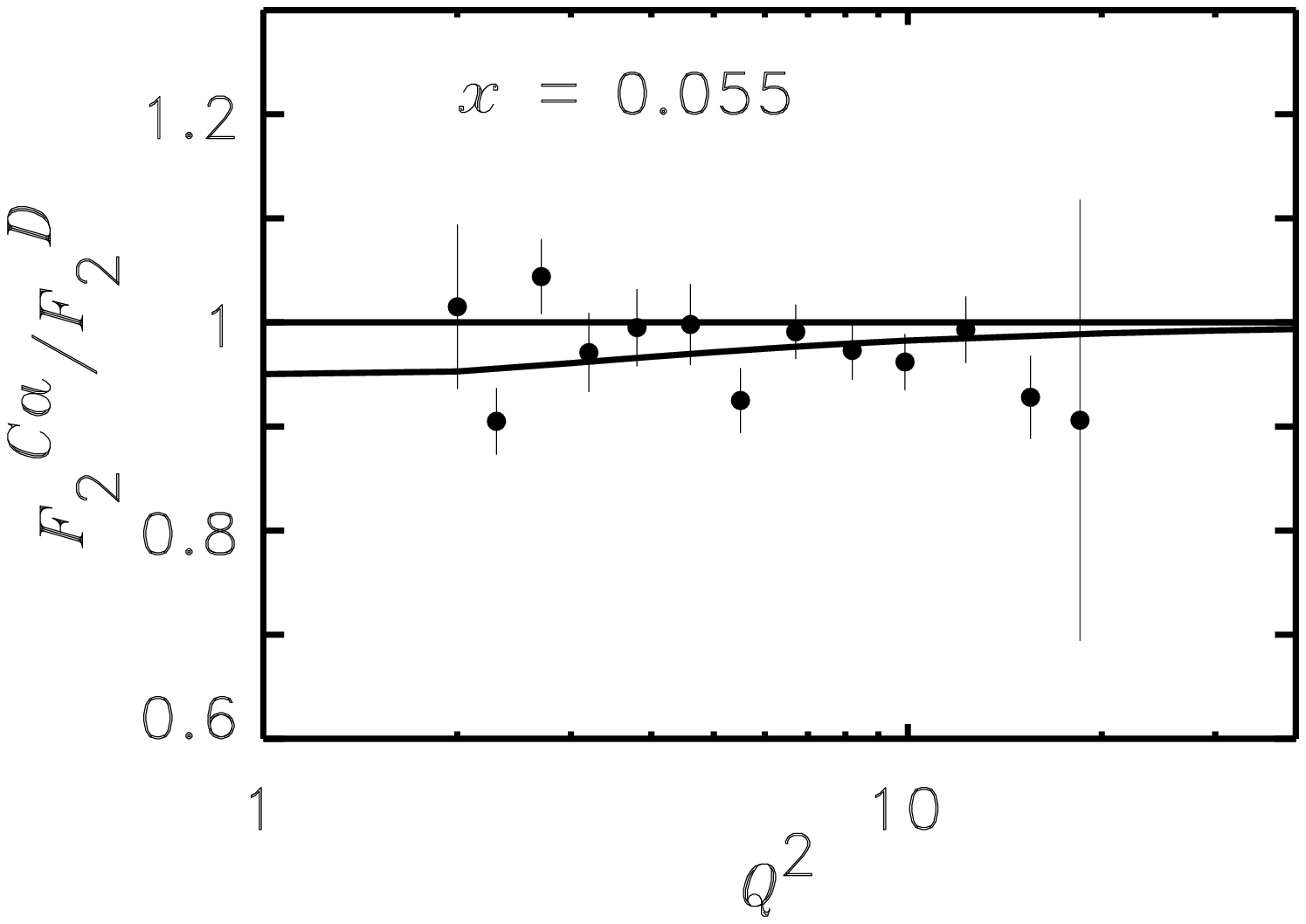,height=7cm}}
\centering{\ \psfig{figure=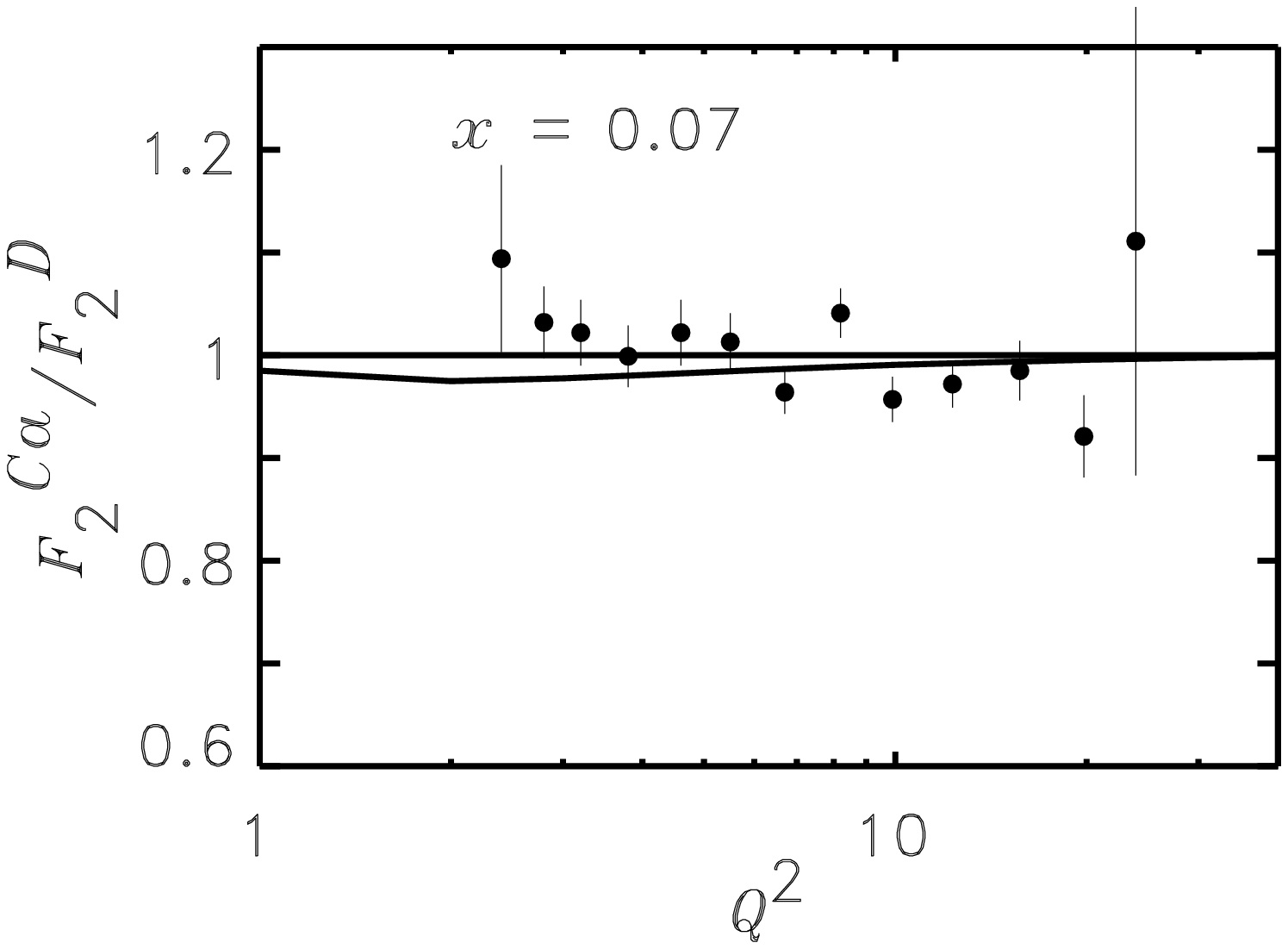,height=7cm}}
\caption{$Q^2$ variation of the $Ca/D$ structure function ratio,
        for $x$ = 0.0085, 0.025, 0.035, 0.045, 0.055 and 0.07,
        compared with the NMC data \protect\cite{N_RE}.}
\end{figure}

\begin{figure}
\centering{\ \psfig{figure=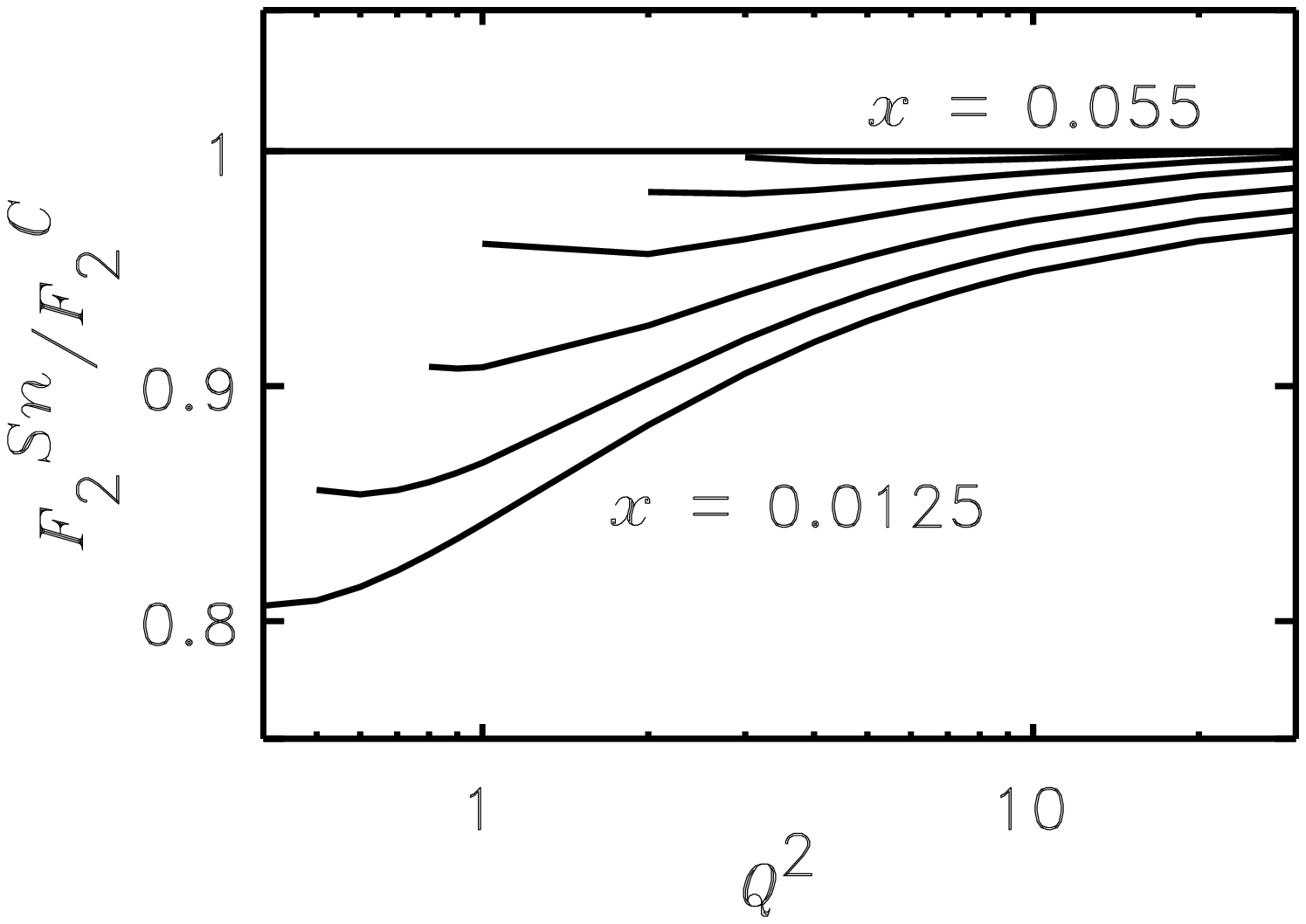,height=9cm}}
\centering{\ \psfig{figure=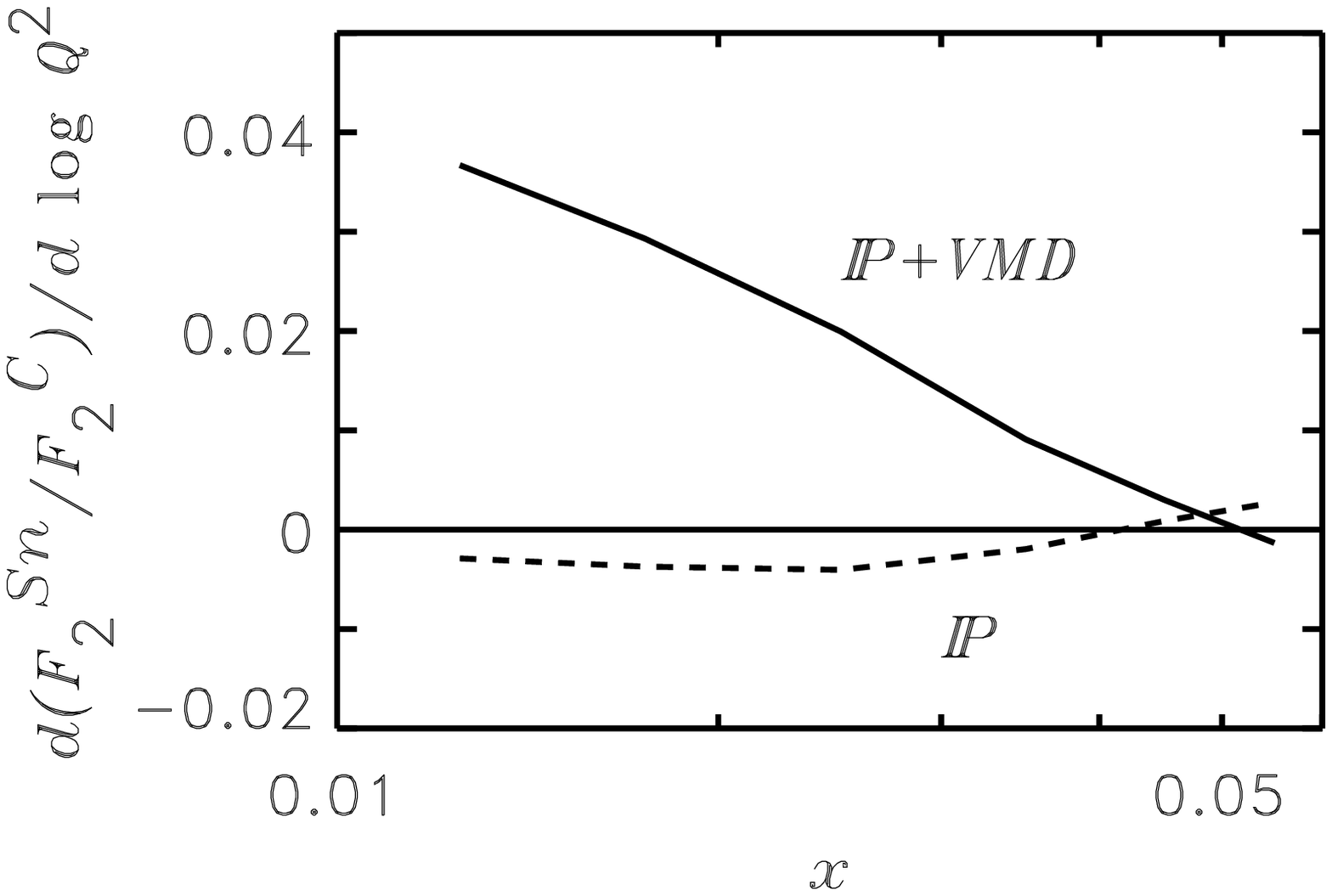,height=9cm}}
\caption{(a) Model prediction for the $Q^2$ dependence of the
        structure function ratio of $Sn$ to $C$,
        for $x$ = 0.0125 (lowest curve), 0.0175, 0.025, 0.035,
        0.045 and 0.055 (highest curve).
         (b) Slope in $\log Q^2$ of the $Sn/C$ ratio as a
        function of $x$ --- solid curve is the full result,
        dashed is the Pomeron contribution only.}
\end{figure}

\begin{figure}
\centering{\ \psfig{figure=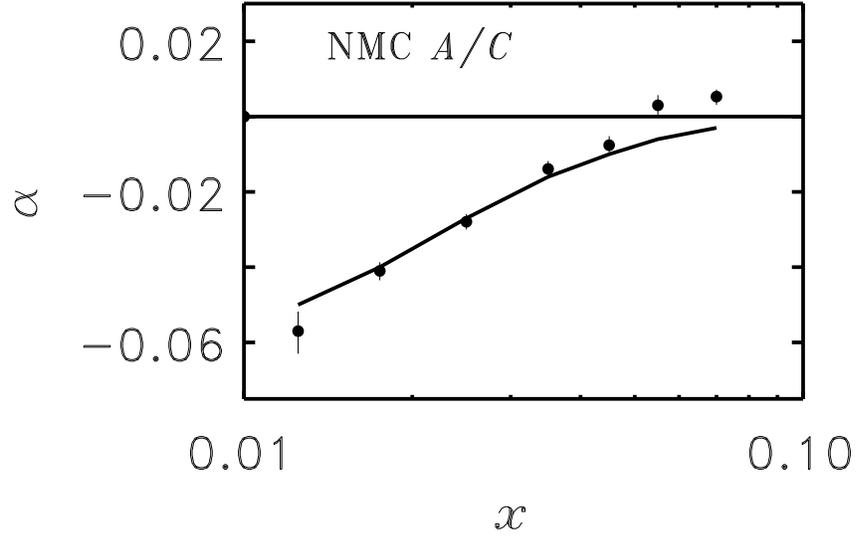,height=9cm}}
\caption{$x$ dependence of the slope $\alpha$ from the structure
        function ratio $F_2^A/F_2^C \propto A^{\alpha}$,
        compared with NMC data on $A = D, Li, Be, Al, Ca, Fe$
        and $Sn$.}
\end{figure}

\begin{figure}
\centering{\ \psfig{figure=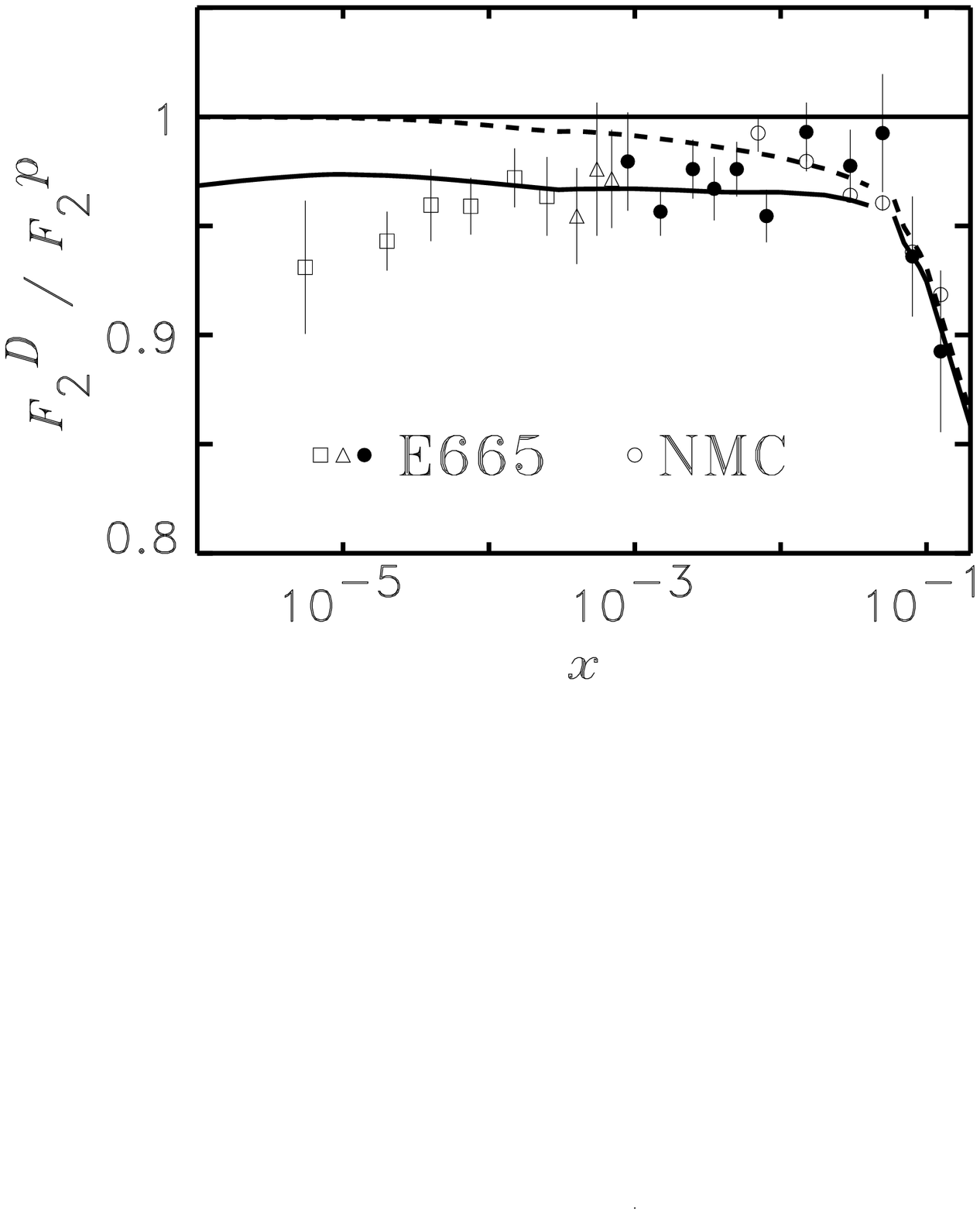,height=9cm}}
\caption{$x$ dependence of the $D/p$ structure function ratio,
        compared with the low-$x$ E665 data \protect\cite{E_D}
        and NMC data \protect\cite{N_D} at larger $x$.
        The dashed curve is the result without any shadowing
        correction.}
\end{figure}

\end{document}